\documentclass{svjour2}                    

\smartqed  
\usepackage{graphicx}
\usepackage{amssymb}
\begin{document}

\title{Flat histogram Monte Carlo simulations of triangulated fixed-connectivity surface models
}

\titlerunning{Flat histogram Monte Carlo simulations of triangulated fixed-connectivity surface models}        

\author{Hiroshi Koibuchi}


\institute{Hiroshi Koibuchi \at
Department of Mechanical and Systems Engineering, Ibaraki National College of Technology, Nakane 866, Hitachinaka, Ibaraki 312-8508, Japan
 \\
              \email{koibuchi@mech.ibaraki-ct.ac.jp}           
}

\date{Received: date / Accepted: date}

\maketitle

\begin{abstract}
Using the Wang-Landau flat histogram Monte Carlo (FHMC) simulation technique, we were able to study two types of triangulated spherical surface models in which the two-dimensional extrinsic curvature energy is assumed in the Hamiltonian. The Gaussian bond potential is also included in the Hamiltonian of the first model, but it is replaced by a hard-wall potential in the second model. The results presented in this paper are in good agreement with the results previously reported by our group. The transition of surface fluctuations and collapsing transition were studied using the canonical Metropolis Monte Carlo simulation technique and were found to be of the first-order. The results obtained in this paper also show that the FHMC technique can be successfully applied to triangulated surface models. It is non-trivial whether the technique is applicable or not to surface models because the simulations are performed on relatively large surfaces.
\keywords{Triangulated surfaces \and Collapsing transition \and Surface fluctuations \and Flat histogram Monte Carlo}
\end{abstract}
\section{Introduction}
Surface models for membranes and strings constructed by Helfrich and Polyakov are described using the notion of two-dimensional differential geometry \cite{HELFRICH-1973,POLYAKOV-NPB1986}. The surface shape in ${\bf R}^3$ is considered to be governed by a curvature Hamiltonian, which is given by an integral over the squared mean curvature or the extrinsic curvature. Consequently, the surface strength is characterized by bending rigidity $b$ \cite{NELSON-SMMS2004,David-TDQGRS-1989,David-SMMS2004,Wiese-PTCP2000,Bowick-PREP2001}. Thus, we understand that the surface collapses and wrinkles in the limit of $b\to 0$ while it swells and becomes smooth in the limit of $b\to\infty$  \cite{KANTOR-NELSON-PRA1987,AMBJORN-NPB1993,WHEATER-JP1994}. Theoretical studies utilizing the renormalization group technique predict that the crumpling transition is continuous \cite{Peliti-Leibler-PRL1985,DavidGuitter-EPL1988,PKN-PRL1988,BKS-PLA2000,Kownacki-Mouhanna-2009PRE}, while density-matrix renormalization group studies on the folding of triangular lattice \cite{NISHIYAMA-PRE-2004} and recent numerical simulations of the model on triangulated surfaces \cite{KD-PRE2002,KOIB-PRE-2004,KOIB-PRE-2005,KOIB-NPB-2006} indicate that the crumpling transition is of the first-order and accompanies the collapsing transition. 

By including certain inhomogeneous components such as cytoskeletal structure or holes in the above-mentioned homogeneous models, we obtain a variety of surface models for numerical studies \cite{KOIB-JSTP-2007,KOIB-PRE-2007-1,KOIB-EPJE-2008}. Lateral diffusion of lipids can also be implemented in the models by the so-called dynamical triangulation technique, which introduces non-uniform coordination numbers $q$ to the triangulated surfaces \cite{KOIB-PRE-2007-2-3}. We should note that such non-uniform $q$ naturally appears in diagrammatic expansions of the matrix integral in the matrix model of $2D$ quantum gravity, where the surface is embedded in the $D\!=\!0$ dimensional space, or equivalently it is not embedded in any external spaces \cite{FGZ-PREP-1995}. 

In those inhomogeneous surface models in ${\bf R}^3$, the transitions separating two neighboring phases are discontinuous \cite{KOIB-JSTP-2007,KOIB-PRE-2007-1,KOIB-EPJE-2008}. Thus, the first-order nature of transitions seems to be a common feature of the shape transformation transitions in the triangulated surface models. Therefore, careful numerical studies are still needed to understand the phase structure of surface models, because first-order transitions are not always easy to analyze numerically.   

In this paper, we study conventional homogeneous surface models on triangulated fixed-connectivity surfaces of sphere topology using the flat histogram Monte Carlo (FHMC) simulation technique of Wang and Landau \cite{Wang-Landau-PRL-2001}. The surfaces are allowed to self-intersect. They are called phantom, or self-intersecting, surfaces.
 
The transitions of homogeneous surface models are already reported to be of the first-order. They were obtained using the canonical Metropolis Monte Carlo (MMC) simulation technique, as mentioned above \cite{KD-PRE2002,KOIB-PRE-2004,KOIB-PRE-2005,KOIB-NPB-2006}. MMC is a simple and reliable technique for studying phase transitions in all statistical mechanical systems. However, MMC is not always an efficient technique for analyzing first-order transitions. The curvature energy of surface models jumps at the first-order transition point. Consequently, the surface configuration can be trapped in one of the minimum energy states if the lattice size increases. The configurations are trapped because MMC is directly defined based on a canonical ensemble, where the configuration of dynamical variables are generated by the Boltzmann weight, which confines the configurations to a narrow energy range. 

In contrast to MMC, FHMC is defined by a random walk in energy space. A large number of studies have been conducted to verify that FHMC is an efficient and reliable technique for phase transitions in spin models \cite{Janke-histogram-2002,Berg-Janke-PRL-2007,SBM-IJMPC-2002,Wang-Landau-PRL-2001,Wang-Landau-2001}. FHMC technique can also be applied to models of polymer chains, in which the phase space is non-compact \cite{TPBinder-JCP2009}, and also to models with more than two energy terms \cite{SRPBinder-JCP2008,Schulz-Binder-PRE2005}. The non-compactness of phase space and the multiple energy terms in those models share the common property with the surface models. In fact, the phase space of surface models is ${\bf R}^3$, which is non-compact and is in sharp contrast to the compact phase space of spin models. Moreover, the Hamiltonian of surface models is always composed of two energy terms: the linear combination of a bond potential $S_1$ and a bending energy $S_2$. However, it is nontrivial whether such a sophisticated technique is useful for studying phase transitions in surface models. In fact, a large scale simulation is necessary to study phase transitions of surface models, while FHMC technique is considered to be problematic on large systems of a spin model \cite{Schulz-Binder-PRE2005}.  

The purpose of this study is two-fold. The first part aims to determine whether FMHC technique can be successfully applied to first-order transitions of surface models. The second purpose is to confirm that the conventional, homogeneous models undergo the first-order transition, which was first assessed by MMC simulation as mentioned above. 

The density of energy $\Omega$ depends on two independent energies, $S_1$ and $S_2$, such that $\Omega\!=\!\Omega(S_1,S_2)$. Therefore, we must replace  $\Omega(S_1,S_2)$ by the single energy density $\Omega(S_2)$, because it seems difficult to obtain $\Omega(S_1,S_2)$ due to the lack of computational speed currently available. Thus, we should check whether this replacement is well defined or not.

\section{Models}\label{model}
The models are defined by the partition function
\begin{equation}
\label{part-func}
Z = \int^\prime \prod_{i=1}^N d X_i \exp\left[ -S(X) \right], 
\end{equation}
where $N$ is the total number of vertices and $S(X)$ is the Hamiltonian. The symbol $\prod_{i=1}^N d X_i$ denotes $3N$-dimensional integration in ${\bf R}^3$. $\int^\prime$ indicates that the integrations are performed such that the surface center is fixed to remove the translational zero mode. The self-avoiding property of the surface is not assumed.

Spherical surfaces in ${\bf R}^3$ are triangulated, and $S(X)$ is defined on them. The triangulated surfaces are constructed from the icosahedron by splitting the edges into $\ell$ pieces of uniform length and dividing the faces into triangles accordingly. The surfaces are identical to those used in \cite{KOIB-PRE-2005}. The total number of vertices is thus given by $N\!=\!10\ell^2\!+\!2$. The total number of bonds $N_B$ and the total number of triangles $N_T$ are given by $N_B\!=\!30\ell^2$ and $N_T\!=\!20\ell^2$, respectively. The coordination number $q$ is $q\!=\!6$ throughout the lattice except at $12$ vertices, which are the vertices of the icosahedron and of $q\!=\!5$.  

The first model denoted by {\it model 1} is defined by the Hamiltonian $S(X)$, which is the linear combination of the Gaussian bond potential $S_1$ and the bending energy $S_2$ with the bending rigidity $b$:
\begin{eqnarray}
\label{model-1}
&&S(X)=S_1+b S_2, \quad S_1=\sum_{(ij)} \left(X_i-X_j\right)^2, \nonumber \\
&&S_2=\sum_{(ij)}\left(1-{\bf n}_i\cdot{\bf n}_j \right),  \qquad({\rm model \; 1}).
\end{eqnarray}
$\sum_{(ij)}$ in $S_1$ denotes the sum over bond $(ij)$ connecting the vertices $i$ and $j$. $\sum_{i,j}$ in $S_2$ is the sum over triangles $i$ and $j$, which share a common bond. The symbol $(X_i\!-\!X_j)^2$ in $S_1$ is the bond length squares between the vertices $i$ and $j$. The symbol ${\bf n}_i$ in $S_2$ denotes a unit normal vector of the triangle $i$. The inner product of the normal vectors ${\bf n}_i\cdot{\bf n}_j$ can also be represented by $\cos \theta_{ij}$, where   $\theta_{ij}$ is the edge angle between two triangles. The unit of $b$ is $kT$, where $k$ is the Boltzmann constant and $T$ is the temperature. 

The second model denoted by {\it model 2} is defined by the linear combination of a hard wall potential $V_{r_0}$ and the bending energy $S_2$ with the bending rigidity $b$ such that
\begin{eqnarray}
\label{model-2}
&&S(X)=V_{r_0}+b S_2, \quad V_{r_0}=\sum_{(ij)} V(|X_i-X_j|), \nonumber \\
&&S_2=\sum_{(ij)}\left(1-{\bf n}_i\cdot{\bf n}_j \right),  \qquad({\rm model \; 2}),
\end{eqnarray}
where $V_{r_0}$ denotes that the potential depends on the parameter $r_0$. The symbol $V(|X_i\!-\!X_j|)$ in $V_{r_0}$ is the potential between the vertices $i$ and $j$ and is defined by  
\begin{equation}
\label{V}
V(|X_i-X_j|)= \left\{
       \begin{array}{@{\,}ll}
       0 & \quad (|X_i-X_j| < r_0), \\  
      \infty & \quad ({\rm otherwise}). 
       \end{array}
       \right. 
\end{equation}
The value of $r_0$ in the right hand side of Eq. (\ref{V}) is fixed at $r_0\!=\!\sqrt{1.1}$. Consequently, we have $\langle \sum (X_i\!-\!X_j)^2 \rangle /N \simeq 3/2$, which is automatically satisfied in model 1 as described below, where the Gaussian bond potential $S_1\!=\!\sum (X_i\!-\!X_j)^2$ is included in the Hamiltonian in place of the hard-wall potential $V_{r_0}$. 

The hard wall potential $V_{r_0}$ makes the mean bond length constant just like the Gaussian bond potential $S_1$, which makes the mean bond length constant in model 1. If it were not for the constraint $|X_i\!-\!X_j| \!<\! r_0$, the size of the surface would grow larger and larger in the MC simulations. Thus the constraint $|X_i\!-\!X_j| \!<\! r_0$ is necessary to make the bond length  well-defined if the Gaussian term $S_1$ is not included in the Hamiltonian. Consequently, model 2 has an additional parameter, $r_0$, which seems to fix a length scale in the model. However,  Ref. \cite{KOIB-PLA-2003-1} shows that the results are not dependent on $r_0$. Therefore, we use $r_0\!=\!\sqrt{1.1}$ in the MC simulations.

We should note that $\langle S_1/N\rangle\!=\!3/2$ is satisfied in model 1. This is understood from the scale invariant property of $Z$ \cite{WHEATER-JP1994}. In fact, by rescaling the integration variable in $Z$ such that $X\to\alpha X$, we obtain $Z(\alpha)\!=\!\alpha^{3(N\!-\!1)}\int^\prime \prod_{i=1}^N d X_i \exp\left[-S(\alpha X) \right]$, where $S(\alpha X)\!=\!\alpha^2 S_1\!+\!bS_2$. The scale invariance of $Z$ indicates that $Z(\alpha)$ is independent of $\alpha$ and, therefore, is represented by $\partial Z(\alpha)/\partial \alpha |_{\alpha=1} \!=\!0$. Thus, we have $\langle S_1/N\rangle\!=\!3(N\!-\!1)/2N\!\simeq\!3/2$. In contrast, $\langle S_1/N\rangle\!=\!3/2$ is not satisfied in model 2 because the partition function $Z$ is not scale invariant in this case. In fact, the scale transformation $X\to\alpha X$ non-trivially changes the potential $V_{r_0}$ because $r_0$ also changes. Therefore, $\partial Z(\alpha)/\partial \alpha |_{\alpha=1} \!=\!0$ is not always satisfied. However, $\langle S_1/N\rangle\!=\!3/2$ is almost satisfied in model 2 as we will see below. This finding implies that the violation of the scale invariance is not very large in model 2.  

$S(X)$ in Eq. (\ref{model-1}) naively has an expression $S(X)\!=\!aS_1\!+\!bS_2$ with a parameter $a$ of the unit $[kT/L^2]$, where $[L]$ is the length unit. By using the above mentioned scale invariance of $Z$, $S(X)$ can also be written as  
$S(X)\!=\!S_1\!+\!bS_2$. In fact, the partition function $Z$ is unchanged by rescaling $X\!\to\!X/\sqrt{a}$, while this rescaling changes $S\!=\!aS_1\!+\!bS_2$ to $S\!=\!S_1\!+\!bS_2$. Thus, we can use this simple expression as the Hamiltonian, where we should always remind ourselves that the coefficient of $S_1$ is assumed to be $a[kT/L^2]\!=\!1$.

\section{Flat Histogram Monte Carlo technique}
\subsection{Histogram and reweighting}
We first describe the histogram technique for model 1 using the terminologies described in \cite{Janke-histogram-2002}. As mentioned in the Introduction, the phase space of the models is ${\bf R}^3$, and the dynamical variables $X(\in {\bf R}^3)$ are continuous. Consequently, the energies $S_1$ and $S_2$ of the models are continuous in contrast to the energy of the Ising spin model. For this reason, the number of states $\Omega (S_1,S_2)$ at energies $S_1$ and $S_2$ should be understood with the symbol $dS_1dS_2$, such that $dS_1dS_2 \Omega (S_1,S_2)$.  Thus, the partition function $Z(b_0)$ in Eq. (\ref{part-func}) can also be described by $\Omega (S_1,S_2)$, such that
\begin{equation}
\label{part-func-2}
Z(b_0) = \int\!\!\!\int dS_1dS_2 \Omega(S_1,S_2)\exp\left[ -(S_1+b_0S_2) \right].
\end{equation}
Let $P_{b_0}$ be $P_{b_0}\propto \Omega (S_1,S_2)\exp\left[ -(S_1+b_0S_2)\right]$. We then obtain $P_{b}\propto P_{b_0}\exp\left[ -(b-b_0)S_2\right]$. Using this, we have the expectation value of a physical quantity $Q$ at $b$ such that  
\begin{equation}
\label{reweighting-1}
\langle Q(b)\rangle=\int\!\!\!\int  dS_1dS_2Q P_b(S_1,S_2)/ \int\!\!\!\int dS_1dS_2  P_b(S_1,S_2).
\end{equation}
This expression is a basic formula for reweighting called a single histogram technique, which allows us to obtain $\langle Q(b)\rangle $ from the Monte Carlo data $P_{b_0}$ at $b_0$. 

The reweighting technique described above has two tasks. The first is to obtain $\Omega (S_1,S_2)$, and the second is to obtain the canonical expectation value by reweighting. The problem lies is performing the first task or obtaining $\Omega (S_1,S_2)$ efficiently. It is well known that the MMC technique is less efficient for evaluating $\Omega $ in a large system because of the exponential dumping of $\Omega$ in the energy space. The multicanonical Monte Carlo simulation (MCMC) technique is considered a dynamic version of the multihistogram reweighting technique \cite{Janke-histogram-2002}. However, it is not apparent whether this technique is applicable to the simulations of large surfaces. In fact, our preliminary study indicates that a flat histogram is barely obtained by MCMC simulations on the surfaces of $N \!\geq\! 5762$ for model 1. By contrast, a flat histogram can be obtained even on relatively large surfaces by the flat-histogram MC technique, which was recently proposed by Wang and Landau. Detailed information on the FHMC technique for surface simulation is described in the following subsection. 

\subsection{Flat histogram Monte Carlo}

The Wang-Landau FHMC technique for surface models, which is simply denoted as FHMC, consists of the following three steps \cite{Janke-histogram-2002}:
\begin{enumerate}
\item[1)] Recursive construction of $\Omega (S_2)$.
\item[2)] A production run with $\Omega (S_2)$, collecting measurements.
\item[3)] Reweighting to extract the canonical expectation of physical quantities.
\end{enumerate}
The first and the second steps are long Monte Carlo runs, but the third step is very short by comparison. The second run (=Step 2) includes thermalization Monte Carlo sweeps (MCS). In the following, we describe each step in detail. 

In FHMC simulations for model 1 and model 2, we use a reduced density of states $\Omega (S_2)$, which is obtained by the replacement
\begin{equation}
\label{Replace_Omega}
\Omega (S_1,S_2) \to \Omega (S_2).
\end{equation}
This replacement is necessary because $\Omega (S_1,S_2)$ is a double histogram and, hence, seems very hard to obtain accurately. This replacement is also well defined if we recall that $S_1$ is almost constant in model 1 due to the scale invariance of the partition function. The fact that $S_1$ is almost constant implies that the role of $S_1$ is to make the mean bond length constant in model 1. Therefore, the potential $S_1$ can be replaced by a Lennard-Jones type potential \cite{KD-PRE2002} or by a hard-wall potential, which is the potential in model 2. Thus, the phase structure of the surface models is primarily dependent on the curvature energy $S_2$. For this reason, the replacement in Eq. (\ref{Replace_Omega}) is considered as reasonable. The histogram $\Omega(S_2)$ is defined by  $S_2$ in the region $S_2^{\rm min}\!<\!S_2\!<\!S_2^{\rm max}$, where $S_2^{\rm min}$ and $S_2^{\rm max}$ are chosen such that the phase transition region is included. We should note that $\Omega(S_2)$ implicitly depends on $S_1$ in model 1 and on $V_{r_0}$ in model 2, although the replacement of Eq. (\ref{Replace_Omega}) is assumed.

In Step 1, $\Omega (S_2)$ is recursively obtained by updating the variables $X$ as follows: the new position $X^\prime_i$ of the vertex $i$ is given by $X^\prime_i\!=\!X_i\!+\!{\it \Delta} X$, where ${\it \Delta X}$ is chosen randomly in a small sphere. The new position $X^\prime_i$ is accepted with the probability ${\rm Min}[1,\Omega(S_2)/\Omega(S_2^\prime)]$, where $S_2$ and $S_2^\prime$ are given by $S_2\!=\!S_2({\rm old})$ and  $S_2^\prime\!=\!S_2({\rm new})$. The radius of the small sphere is chosen so that the acceptance rate $r_X$ for $X^\prime$ is approximately $r_X\!=\!50\%$. However, $r_X$ varies during the simulations in Steps 1 and 2 in contrast to the MMC case, where $r_X$ remains almost constant. The constraint $3/2\!-\!{\it \Delta} \!<\!S_1({\rm new})/N\!<\! 3/2$ is also imposed on $X^\prime_i$ only in model 1, while the constraint $V$ in Eq. (\ref{V}) is imposed on  $X^\prime_i$ in model 2. At the beginning of the simulation of Step 1, MMC simulations are performed as a thermalization MCS to make the variables $X$ satisfy the conditions $3/2\!-\!{\it \Delta} \!<\!S_1({\rm new})/N\!<\! 3/2$ and $S_2^{\rm min}\!<\!S_2\!<\!S_2^{\rm max}$. The parameter ${\it \Delta}$ is fixed to ${\it \Delta}\!=\!0.05$ in model 1. This constraint for $S_1({\rm new})/N$ is imposed on the new variable $X^\prime_i$ so that the replacement of Eq. (\ref{Replace_Omega}) becomes well defined, as mentioned above.

The histogram $\Omega(S_2)$ is updated such that
\begin{equation}
\label{Omega}
\Omega(S_2) \to f \,\Omega(S_2)
\end{equation}
with the initial values $\Omega(S_2)\!=\!1$ and $ f=\exp(1)$. The update of $\Omega(S_2)$ in Eq. (\ref{Omega}) is performed independently, whether $X^\prime$ is accepted or not, in every update of $X$. The formula $\log[\Omega(S_2)] \to \log(f)+\log[\Omega(S_2)] $ is used in the simulations. The multiplicative factor $f$ in Eq. (\ref{Omega}) is also redefined such that
\begin{equation}
\label{m-factor}
f \to \sqrt{f} 
\end{equation}
only when a given condition is satisfied. To perform this redefinition of $f$, we check whether the energy histogram $H(S_2)$ is sufficiently flat or not at every $10^5$ MCS by the condition $H(S_2)\!<\!\epsilon \bar{H}$, where $\bar{H}$ is the mean value of $H(S_2)$ and $\epsilon\!=\!0.9$. Once this flatness condition for $H(S_2)$ is satisfied, the histogram $H(S_2)$ is reset to zero for all $S_2$. This procedure is repeated while $f\!>\!1+10^{-8}$ in Step 1. The total number of ${\rm MCS}_1$ for the recursion is not an input parameter, but an output datum, and its value depends mainly on the conditions of $f$ and $H(S_2)$. The parameters $S_2^{\rm min}$, $S_2^{\rm max}$, and ${\it \Delta} s$ should be provided as input data for the recursion simulations so that the histograms $\Omega(S_2)$ and $H(S_2)$ are defined at $S_2^{\rm min}$, $S_2^{\rm min}\!+\!{\it \Delta} s$,  $S_2^{\rm min}\!+2\!{\it \Delta} s$, $\cdots$, $S_2^{\rm max}\!-\!{\it \Delta} s$. Let $N_H$ be the total number of these energy points. Then, the energy step ${\it \Delta} s$ is given by ${\it \Delta} s\!=\!(S_2^{\rm max}\!-\!S_2^{\rm min})/N_H$.

Step 2 is the production run, which is performed using the histogram $\Omega (S_2)$ and is simply described as follows:  the new position $X^\prime_i$ of the vertex $i$ is accepted with the probability ${\rm Min}[1,\exp(-{\it \Delta} S_1) \Omega(S_2)/\Omega(S_2^\prime)]$, where  ${\it \Delta} S_1\!=\!S_1({\rm new})\!-\!S_1({\rm old})$, and $\Omega(S_2)\!=\!\Omega[S_2({\rm old})]$, $\Omega(S_2^\prime)\!=\!\Omega[S_2({\rm new})]$. The factor $\exp(-{\it \Delta} S_1)$ is set to 1 in model 2. Because $\Omega (S_2)$ is defined in the region $S_2^{\rm min}\!<\!S_2\!<\!S_2^{\rm max}$, the new position $X^\prime_i$ is limited so that $S_2^{\rm min}\!<\!S_2^\prime\!<\!S_2^{\rm max}$ just like in Step 1. In Step 2, no constraint is imposed on the bond length in model 1, while the potential $V$ in Eq. (\ref{model-2}) imposes its constraint on the bond length in model 2. In both models, $S_1$ is expected to be almost constant, such that $S_1/N\!=\!3/2$. The thermalization MCS is fixed to $2\times 10^7\sim $  $5\times 10^7$ in both models. A sufficiently large number of  ${\rm MCS}_2$ for production runs is performed after the thermalization MCS, where ${\rm MCS}_2$ is an input parameter in contrast to ${\rm MCS}_1$  in Step 1. The measurements are performed every 500 MCS in both models. The acceptance rate $r_X$ varies during the simulations in Step 2, just like in Step 1 mentioned above.

Table \ref{table-1} shows the parameters, including $S_2^{\rm min}$ and $S_2^{\rm max}$. $N_H$ is the total number of energy points for the histograms of $\Omega(S_2)$ and $H(S_2)$.  ${\rm MCS}_1$ and  ${\rm MCS}_2$ are the total number of MCSs performed in Step 1 and Step 2, respectively.

\begin{table}[hbt]
\caption{ The parameters used in the FHMC of model 1 and model 2. These are all input parameters, excluding ${\rm MCS}_1$. }
\label{table-1}
\begin{center}
 \begin{tabular}{cccccccc}
model & $N$ & $S_2^{\rm min}/N_B$ & $S_2^{\rm max}/N_B$ & $N_H$ & ${\rm MCS}_1(\times10^8)$  &  ${\rm MCS}_2(\times10^8)$   \\
 \hline
  1  & 2562  & 0.35 & 0.47  & 1000  & $1.8$  & $9$      \\
  1  & 5762  & 0.35 & 0.5   & 2500  & $11.2$ & $17.8$   \\
  1  & 7292  & 0.35 & 0.5   & 3300  & $15.3$ & $27.3$   \\
  1  & 10242 & 0.35 & 0.5   & 4500  & $18.6$ & $49$     \\
  1  & 15212 & 0.35 & 0.5   & 6700  & $44.5$ & $50$     \\
 \hline
  2  & 2562  & 0.34 & 0.49  & 1000  & $6.6$  & $9$      \\
  2  & 4842  & 0.35 & 0.49  & 1500  & $16.8$ & $21.6$   \\
  2  & 8412  & 0.35 & 0.51  & 2500  & $43.2$ & $37.2$   \\
  2  & 16812 & 0.35 & 0.51  & 5550  & $45.3$ & $108$     \\
 \hline
 \end{tabular} 
\end{center}
\end{table}

Step 3 is performed by using a technique that is analogous to the multi-histogram reweighting technique; however, the recursion is not necessary because of the single Monte Carlo simulation performed in Step 2. The canonical expectation value of a physical quantity $Q(b)$ is given by
\begin{equation}
\label{mean_value}
\langle Q(b)\rangle ={ \sum_{S_2}\sum_{Q} Q\,h(S_2,Q)\Omega(S_2)\exp(-bS_2) \over \sum_{S_2}\sum_{Q} h(S_2,Q)\Omega(S_2)\exp(-bS_2)},
\end{equation}
where $h(S_2,Q)$ is the histogram of the data obtained in Step 2. This formula Eq. (\ref{mean_value}) corresponds to Eq. (\ref{reweighting-1}), where $b_0$ is actually set to zero. We should note that the sum over $S_1$ in the denominator/numerator is dropped from Eq. (\ref{mean_value}) in model 1. This is because of the replacement of Eq. (\ref{Replace_Omega}).  A normalization constant can be included in the exponential factor in both the denominator and the numerator in Eq. (\ref{mean_value}).

We use a random number called Mersenne Twister \cite{Matsumoto-Nishimura-1998}. A sequence of uniform random numbers is used for a 3-dimensional move of the vertices $X$ and for FHMC accept/reject decisions in the update of $X$.
\section{Results}
The mean square size $X^2$ is defined by
\begin{equation}
\label{X2}
X^2={1\over N} \sum_i \left( X_i -\bar X \right)^2,\qquad \bar X ={1\over N} \sum_i X_i,
\end{equation}
where $\sum_i$ denotes the sum over vertices $i$, and $\bar X$ is the center of the surface.  $X^2$ is identical to the radius squares if the surface becomes smooth and spherical, while it becomes $X^2\!\to\!0$ if the surface collapses. Therefore, $X^2$ can reflect the surface size.    

\begin{figure}[hbt]
\centering
\includegraphics[width=9.5cm]{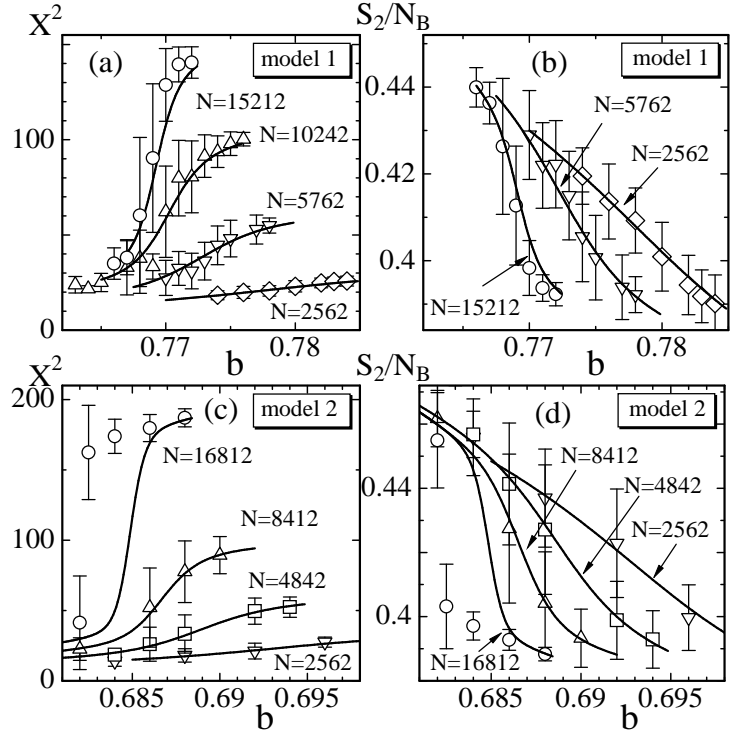}
  \caption{(a) The mean square size $X^2$ vs. $b$ of model 1, (b) the bending energy $S_2/N_B$ vs. $b$ of model 1, (c) $X^2$ vs. $b$ of model 2, and (d) $S_2/N_B$ vs. $b$ of model 2. The solid curves are the results of FHMC, while the symbols ($\bigcirc$, $\bigtriangleup $, $\cdots$) are those of MMC in \cite{KOIB-PRE-2005,KOIB-NPB-2006}.}
\label{fig-1}
\end{figure}
Figures \ref{fig-1}(a) and \ref{fig-1}(b) show $X^2$ vs. $b$ and the bending energy $S_2/N_B$ vs. $b$ of model 1, respectively. $N_B(=\!3N\!-\!6)$ is the total number of bonds. The solid curves are the results of FHMC, while the symbols such as the circle, triangle, etc., are those of MMC, which are identical to the results reported in \cite{KOIB-PRE-2005}. The error bars for the symbols in this paper are standard errors, which are obtained by binning analysis. From these symbols, we can directly see whether the results of FHMC coincide with those of MMC. For this reason, MMC data from \cite{KOIB-PRE-2005} are presented in this paper. Figures \ref{fig-1}(c) and \ref{fig-1}(d) show the results of model 2. The symbols are the results of MMC reported in \cite{KOIB-NPB-2006}. In the case of model 1, we see that the FHMC results are in good agreement with MMC in both $X^2$ and $S_2/N_B$ at $N\!\leq\!10242$. On the $N\!=\!15212$ surface, the results are only slightly different from each other. The FHMC results of model 2 are also in good agreement with those of MMC at $N\!\leq\!8412$. A deviation can be seen in the data obtained on the largest surface of $N\!=\!16812$. We have no definite explanation for these deviations. However, the total number of MMCs was not sufficiently large, at least on the largest surfaces in model 1 \cite{KOIB-PRE-2005} and in model 2 \cite{KOIB-NPB-2006}. Note also that the FHMC results seem to be dependent on the choice of $S_2^{\rm min}/N_B$ and $S_2^{\rm max}/N_B$ if the range is narrow. A wide range is better for accuracy in the obtained results. However, the simulation time becomes longer and longer with increasing range on such large surfaces.    

\begin{figure}[hbt]
\centering
\includegraphics[width=9.5cm]{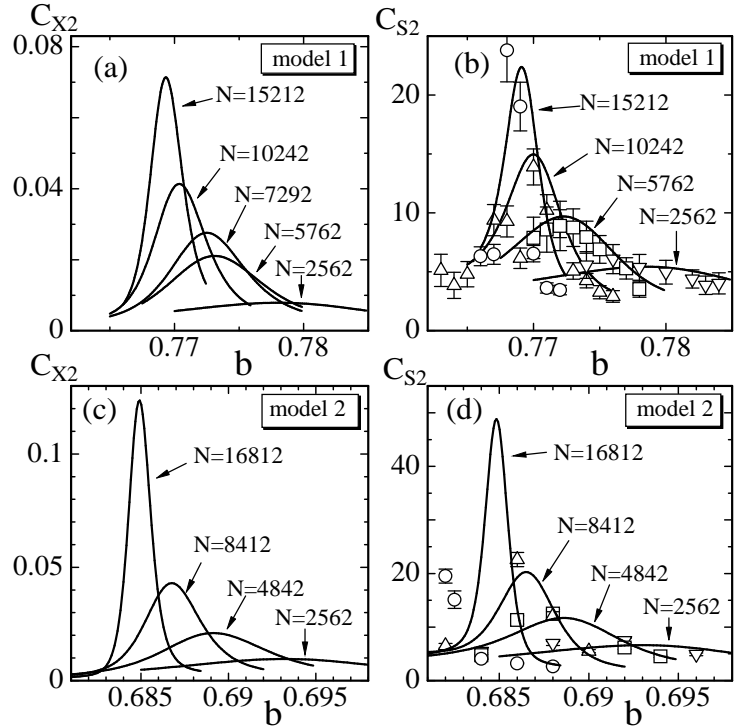}
\caption{(a) The variance $C_{X^2}$ vs. $b$ of model 1, (b) the specific heat $C_{S_2}$ vs. $b$ of model 1, (c) $C_{X^2}$ vs. $b$ of model 2, and (d) $C_{S_2}$ vs. $b$ of model 2. The solid curves are the results obtained by FHMC, while the symbols are those of MMC.}
 \label{fig-2}
\end{figure}
The variance $C_{X^2}$ of $X^2$ is defined by 
\begin{equation}
\label{CX2}
C_{X^2}={1\over N} \langle \left(X^2-\langle X^2\rangle \right)^2 \rangle.
\end{equation}
This variance $C_{X^2}$ reflects the fluctuations of $X^2$ around the mean values. The specific heat $C_{S_2}$ of $S_2$ is defined by 
\begin{equation}
\label{CS2}
 C_{S_2} ={b^2\over N} \langle  \left( S_2 \!-\! \langle S_2 \rangle\right)^2\rangle. 
\end{equation}
This $C_{S_2}$ also reflects the fluctuations of the bending energy. 

Figure \ref{fig-2}(a) shows $C_{X^2}$ vs. $b$ of model 1, while Fig. \ref{fig-2}(b) shows $C_{S_2}$ vs. $b$ of model 1. The results of model 2 are also shown in Figs. \ref{fig-2}(c) and \ref{fig-2}(d). The solid curves in the figures are the results obtained by FHMC, while the symbols are those obtained by MMC, as presented in Figs. \ref{fig-1}(a)--\ref{fig-1}(d). In Fig. \ref{fig-2}(b), the solid curves of $C_{S_2}$ almost agree with the symbols. However, there is a small difference between the two results at $N\!=\!15212$. The specific heat $C_{S_2}$ obtained by FHMC is not always in good agreement with those obtained  by MMC, as can be seen in model 2 in Fig. \ref{fig-2}(d). We feel that the reason for this deviation is mainly due to a lack of statistics in MMC simulations in \cite{KOIB-PRE-2005,KOIB-NPB-2006}. The MMC data of $C_{X^2}$ in \cite{KOIB-PRE-2005,KOIB-NPB-2006} are not shown in Figs. \ref{fig-2}(a) and  \ref{fig-2}(c) because the deviation between the MMC data and the solid lines is relatively large compared to those in $C_{S_2}$. The variance $C_{X^2}$ and the specific heat $C_{S_2}$ are not so easy to obtain accurately. However, we observe that the peaks of $C_{X^2}$ and $C_{S_2}$ grow with increasing $N$. Therefore, our expectation is that the phase transitions are reconfirmed by FHMC simulations. 

\begin{figure}[hbt]
\centering
\includegraphics[width=9.5cm]{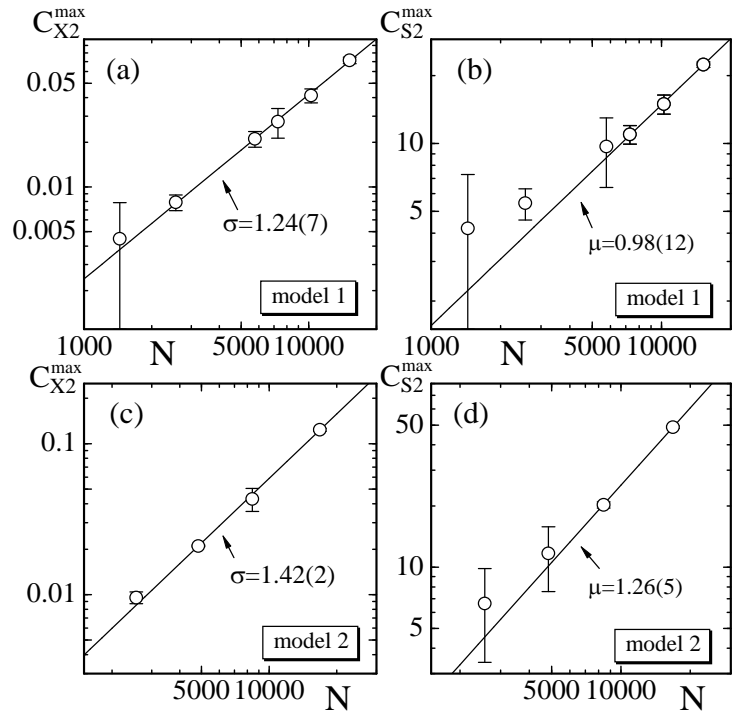}
\caption{Log-log plots of the peak values $C_{X^2}^{\rm max}$ vs. $N$ and $C_{S_2}^{\rm max}$  vs. $N$ of model 1 and model 2. The error bars for the data are standard errors. The data and the errors were obtained by FHMC. The straight lines are drawn by fitting the data to Eq. (\ref{scaling_12}). The largest three data points are used in the fitting in the cases of (b) and (d).}
 \label{fig-3}
\end{figure}
The peak values $C_{X^2}^{\rm max}$ of model 1 are shown as symbols in Fig. \ref{fig-3}(a) against $N$ in a log-log scale. The peaks $C_{S_2}^{\rm max}$ of model 1 are shown in Fig. \ref{fig-3}(b). The error bars drawn on the symbols denote the standard errors of $C_{X^2}^{\rm max}$ and $C_{S_2}^{\rm max}$. The straight lines in Figs. \ref{fig-3}(a) and \ref{fig-3}(b) are drawn by fitting the data to 
\begin{equation}
\label{scaling_12}
C_{X^2}^{\rm max} \sim N^\sigma,\qquad C_{S_2}^{\rm max} \sim N^\mu,
\end{equation}
where $\sigma$ and $\mu$ are critical exponents. The fitting of $C_{X^2}^{\rm max}$ in Fig. \ref{fig-3}(b) is performed by using the largest three data points. Figures \ref{fig-3}(c) and \ref{fig-3}(d) are the results obtained by model 2. The symbols and the errors are simulation data. The straight lines are drawn by fitting the data. The fitted line in Fig. \ref{fig-3}(d) is obtained using the largest three data points. Thus we have
\begin{eqnarray}
\label{exponents}
&&\sigma = 1.24\pm0.07, \quad \mu = 0.98\pm0.12,\quad ({\rm model \; 1}), \nonumber \\
&&\sigma = 1.42\pm0.02, \quad \mu = 1.26\pm0.05,\quad ({\rm model \; 2}). 
\end{eqnarray}
The exponent $\mu $ of model 1 is almost identical to $\mu\!=\!1 $. Therefore, the transition of surface fluctuations in model 1 is considered to be of the first order. We also see that the remaining exponents are consistent with a first-order collapsing transition and a first-order transition of surface fluctuations, although they are slightly larger than $1$. These results are consistent with the previous conclusions described in \cite{KOIB-PRE-2005,KOIB-NPB-2006}.  
We should emphasize that the symbols and the errors in Figs. \ref{fig-3}(a)--\ref{fig-3}(d) are the ones obtained by the FHMC technique in contrast to those shown in the previous figures, such as Fig. \ref{fig-1} and Fig. \ref{fig-2}, where the symbols correspond to MMC simulation data in \cite{KOIB-PRE-2005,KOIB-NPB-2006}.

\begin{figure}[hbt]
\centering
\includegraphics[width=9.5cm]{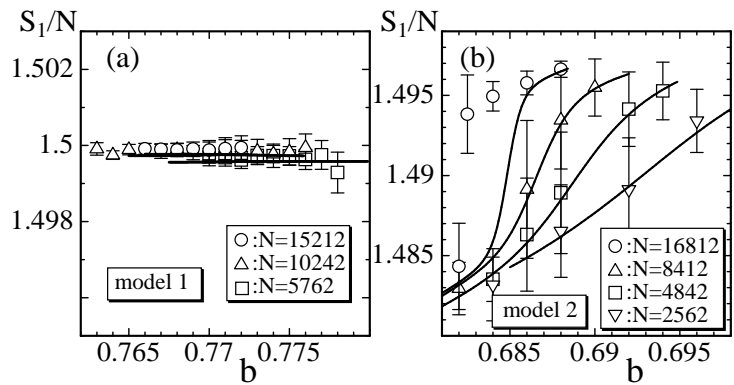}
 \caption{The Gaussian bond potential $S_1/N$ vs. $b$ of (a) model 1 and (b) model 2. The solid curves are obtained by FHMC, while the symbols are MMC data. }
\label{fig-4}
\end{figure}
Finally in this subsection, we show the Gaussian bond potential $S_1/N$ vs. $b$ in Figs. \ref{fig-4}(a) and \ref{fig-4}(b). The solid curves represent FHMC results, while the symbols are the results of MMC, just as presented in Figs. \ref{fig-1} and \ref{fig-2}. As described in Section \ref{model}, $S_1/N$ in model 1 is expected to be $S_1/N\!=\!3/2$. We see from Fig. \ref{fig-4}(a) that this expectation is satisfied just like the MMC data. The solid curves of $S_1/N$ of model 2 in Fig. \ref{fig-4}(b) are also consistent to the results of MMC. We should note that the discontinuity seen in $S_1/N$ reflects a discontinuous change of bond length in model 2. Therefore, the phase transition in model 2 is considered to be accompanied by a structural change of its surface, although the discontinuity is very small compared to the value of $S_1/N$. The fact that the discontinuity in $S_1/N$ is very small compared to $S_1/N$ itself indicates that the violation of scale invariance is small in model 2, as mentioned in Section 2. 

\section{Summary and conclusions}
In this paper, we studied two types of triangulated spherical surface models using the flat histogram Monte Carlo (FHMC) simulation technique of Wang and Landau. The surface models have long been studied numerically, mainly using Metropolis Monte Carlo (MMC) simulations. It has been recently reported that the surface fluctuations in these models undergo a first-order transition. We aimed to confirm that the transition is of the first order and to verify that the FHMC technique can be applied to study phase transitions of surface models.

It is nontrivial whether FHMC can be used to study surface models because large-scale simulations are necessary to study the phase structure of surfaces. It was pointed out in  \cite{Schulz-Binder-PRE2005} that the FHMC technique can not always be successfully applied to large-scale simulations of spin models. On the other hand, the variables $X$ of the surface models are updated by a random walk in energy space in the FHMC technique. Therefore, the FHMC technique is considered to have an advantage in studying first-order phase transitions, as compared to the MMC technique. However, the surface simulations by FHMC remained to be studied. 

The FHMC simulations performed in this paper consist of three steps: the first is a random walk in energy space to obtain the density of states $\Omega(S_2)$ recursively. The second is to collect measurements using $\Omega(S_2)$. The final step is to obtain canonical expectations using a reweighting technique. The first two steps are long MC runs, each of which is a single MC run. We need many parameters to start FHMC simulations. Short MMC simulations are required to obtain these parameters before starting FHMC simulations.     

The results obtained by FHMC simulations were compared to those reported in \cite{KOIB-PRE-2005,KOIB-NPB-2006}, which were obtained by MMC simulations.  FHMC results were found to be in good agreement with MMC results in both model 1 and model 2, excluding the data obtained for the largest surfaces in both models. Although the variances $C_{X^2}$ and $C_{S_2}$ obtained by FHMC simulations are slightly different from those previously obtained by MMC simulations on the largest surfaces, the order of the transitions is not influenced. The reason for these differences seems to be due to the lack of statistics in the previous MMC simulations, at least on the largest surfaces. No problem is observed in FHMC simulations concerning the system size at least on the surfaces upto $N=15000\!\sim\! 17000$. Thus, we conclude that the first-order nature of the transitions is reconfirmed by the FHMC technique and, consequently, that the FHMC technique can be successfully applied to study the phase structure of surface models.




\end{document}